\documentclass[aps,10pt,pra,twocolumn,groupedaddress, floatfix, superscriptaddress,showpacs, showkeys,amsfonts]{revtex4-2}
\usepackage{amsfonts}
\usepackage{amsmath}
\usepackage{amssymb, setspace}
\usepackage{graphics,graphicx}
\graphicspath{{./}}
\usepackage{epsf}
\usepackage{subfigure}
\usepackage[%
  colorlinks=true,
  urlcolor=blue,
  linkcolor=blue,
  citecolor=blue
]{hyperref}
\usepackage[all]{hypcap}
\usepackage{color}
\usepackage[utf8]{inputenc}
\usepackage{dsfont}
\usepackage{enumitem}
\usepackage[utf8]{inputenc}

\usepackage{comment}
\usepackage{orcidlink}
\usepackage[all]{nowidow}
\usepackage{natbib}
\usepackage{braket}
\usepackage{physics}

\usepackage[normalem]{ulem}
\usepackage{xcolor}

\newcommand{\be}{\begin{equation}}
\newcommand{\ee}{\end{equation}}
\newcommand{\bea}{\begin{eqnarray}}
\newcommand{\eea}{\end{eqnarray}}
\newcommand{\eq}[1]{Eq.~\eqref{#1}}

\newcommand{\eqss}[2]{Eqs.~\eqref{#1}-\eqref{#2}}
\newcommand{\seq}[1]{Sec.~\ref{#1}}

\newcommand{\app}[1]{App.~\ref{#1}}
\newcommand{\fig}[1]{Fig.~\ref{#1}}
\newcommand{\figs}[2]{Figs.~\ref{#1}-\ref{#2}}
\newcommand{\bem}{\begin{multline}}
\newcommand{\eem}{\end{multline}}

\newcommand\identity{1\kern-0.25em\text{l}}

\newcommand{\checksym}{{\bullet}}

\newcommand{\ibmil}{IBM Quantum, IBM Research -- Israel, Mount Carmel, Haifa 31905, Israel}

\begin{document}

\title{Low-weight quantum syndrome errors in belief propagation decoding}

\author{Haggai Landa}
\email{haggaila@gmail.com}
\affiliation{\ibmil}

\begin{abstract}
    We describe an empirical approach to identify low-weight combinations of columns of the decoding matrices of a quantum circuit-level noise model, for which belief-propagation (BP) algorithms converge possibly very slowly. Focusing on the logical-idle syndrome cycle of the low-density parity check gross code, we identify criteria providing a characterization of the Tanner subgraph of such low-weight error syndromes. We analyze the dynamics of iterations when BP is used to decode weight-four and weight-five errors, finding statistics akin to exponential activation in the presence of noise or escape from chaotic phase-space domains. We study how BP convergence improves when adding to the decoding matrix relevant combinations of fault columns, and show that the suggested decoder amendment can result in the reduction of both logical errors and decoding time.
\end{abstract}

\maketitle

\section{Introduction and definition of the problem of low-weight errors}\label{sec:define}

Elements of quantum error correction are being realized with various platforms \cite{andersen2020repeated, satzinger2021realizing, sundaresan2023demonstrating, PhysRevX.11.041058,krinner2022realizing, postler2022demonstration,google2023suppressing, gupta2024encoding, bluvstein2023logical, google2025quantum, putterman2025hardware, reichardt2024logical,dasu2026computing}, and architectures for fault-tolerant computations are being designed \cite{litinski2019game, schwerdt2024scalable,ismail2025transversal,gidney2025factor}. Recently, attention has been focused on low-density parity check (LDPC) codes, in particular the family of bivariate-bicycle codes of which a notable representative is the gross code \cite{bravyi2024high}, with a promising fault-tolerant architecture \cite{yoder2025tour}, building on earlier research \cite{gottesman2014fault, breuckmann2021quantum,cohen2022low,cross2024improved, williamson2024low, swaroop2024universal,cowtan2025parallel,zhang2025time}.

In typical physical embeddings of stabilizer codes, cycles of the error correction circuit terminate by measuring the stabilizer parities using check qubits, which constitute syndromes that are fed to a decoder. With LDPC codes and the circuit-level noise model, the decoder is often based on a linearization of the noise dynamics, searching for the (estimated) most likely linear combination of the syndromes of single circuit faults (one-gate errors).

A commonly used algorithm to decode the faults from the syndromes is belief propagation (BP), which, however, is not sufficient on its own; if BP does not converge, something has to be done. The traditional approach is to invoke a secondary decoder (with a significantly higher algorithmic complexity), for example the ordered statistics decoding (OSD) algorithm \cite{roffe_decoding_2020, Roffe_LDPC_Python_tools_2022,bravyi2024high,feedforward,qec_flip_repo}. A recent advancement and much more efficient is Relay-BP \cite{muller2025improved,relaybp}, which uses a combination of improvements to BP, relying on randomizations and memory to explore the decoding phase space. Relay-BP is very successful in general decoding settings \cite{muller2025improved}, and in particular is suitable in the current context with complex dynamics determined by many connections in the Tanner graph of the decoding matrices, as we show below.
Reducing the real-time decoding computational limits is therefore an active area of research from both hardware and theory perspectives \cite{ liyanage2023scalable,bombin2023modular, caune2024demonstrating, gong2024toward,chen2025improved, jones2024improved,kaufmann2025blockbp, demarti2024decoding, higgott2025sparse, PhysRevLett.133.240602,ott2025decision, beni2025tesseract,stambler2023addressing,feedforward,maurer2025real,beverland2025fail,maurya2025fpga,jacoby2026stairway,strikis2026high}.

For the gross code (for which $[[n,k,d]]=[[144,12,12]]$, with $n$ data qubits, $k$ logical qubits, and code distance $d$) and considering the idle logical circuit in \cite{bravyi2024high}, the circuit-level distance with the standard noise model obeys $d_{\rm circ} \le 10$, where $d_{\rm circ}$ is defined to be the minimum
number of faulty gates in a syndrome cycle that can result in a logical error but a null syndrome. Numerical simulations indicate that this is a tight bound, and hence it is expected that within the same noise model any syndrome with no more than $d_{\rm corr} \equiv \lfloor (d_{\rm circ} -1) / 2 \rfloor = 4$ faulty operations can be decoded and corrected without error.

\begin{figure}[t!]
\centering
\includegraphics[width=0.48\textwidth]
{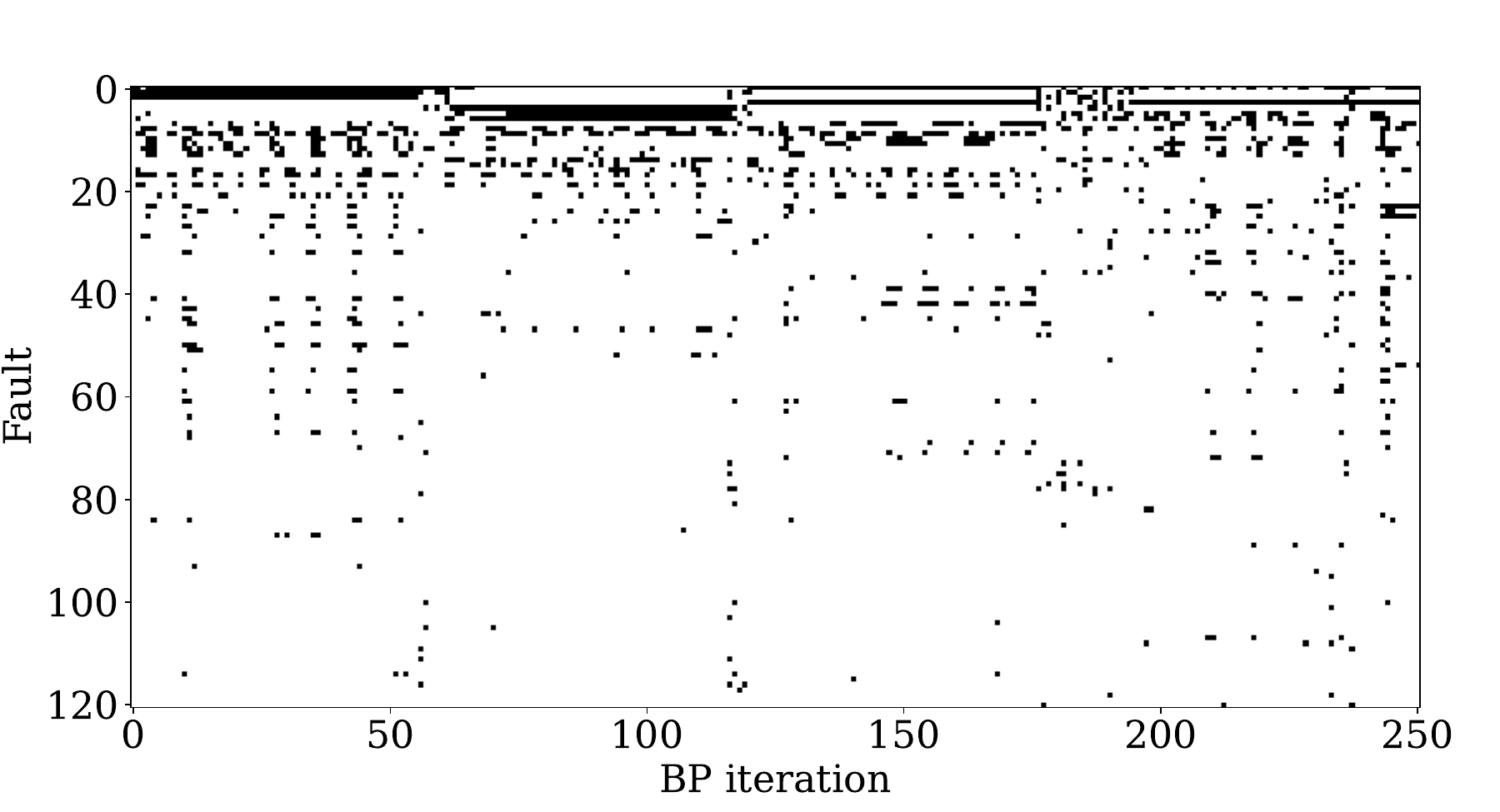}
\caption{The ``hard decision'' result bit of part of the faults (variable nodes) at the end of part of the iterations in the middle of a Relay-BP execution. Each row corresponds to a single fault column in the decoding matrix, and the decoder attempts to decide whether it contributed to the measured syndrome, where a black (white) pixel indicates yes (no).
In this example, a low-weight error syndrome does not converge to a correct result even with 200 relay legs (and over 12,000 BP iterations). There are 60 iterations per relay leg, with a bit more than four legs shown. The faults are ordered by their total (integrated) brightness along all iterations, and only the first 121 from the top are shown. For comparison, a quickly converging weight-four error is presented in \fig{fig:success}.
\label{fig:fault}}
\end{figure}

However, it is found that there are syndromes formed by combinations of four faulty gates whose decoding by BP-based algorithms (Relay-BP and BP-OSD) either converges very slowly, fails to converge, or results in logical errors. Figure \ref{fig:fault} shows an example of a low-weight error syndrome that does not converge within a very long execution of Relay-BP. It can be contrasted with \fig{fig:success} in \app{sec:app}, showing an error converging after 54 BP iterations -- which is still much longer than typical low-weight errors, as detailed below. Examining the decoding dynamics shows that about 10 to 20 candidate faults (Tanner-graph variable nodes) are significantly involved in the dynamics (and many more take part stochastically), and the correct ones comprising the error are not necessarily among those. In this sense the dynamics of the low-weight errors studied here seem to be more complex than typical trapping sets studied in the literature \cite{ivkovic2016eliminating,chytas2025enhanced}, and we elaborate on their characteristics in the following.

In the general case, low-weight errors, syndromes that are combinations of a number of circuit-level faults equal to or less than $d_{\rm corr}$, are important to identify since they play the dominant role in defining an ``error floor'' for the logical error rate, dependent on the code, syndrome measurement circuit, noise model and decoding algorithm and parameters.
In the following sections we focus on the idle cycle of the gross code with separate $X$ and $Z$ decoding, and in \seq{sec:identify} we describe criteria for identifying weight-four errors in this cycle (lower-weight syndromes were not found to manifest logical errors). In \seq{sec:dynamics} we analyze the dynamics of decoding of those faults, and in \seq{sec:resolve} we study the efficacy of an approach to amend the decoding matrices with composite fault columns. We find that this can significantly eliminate decoding issues caused by the identified low-weight errors.

\begin{figure}[t!]
\centering
\includegraphics[width=0.46\textwidth]
{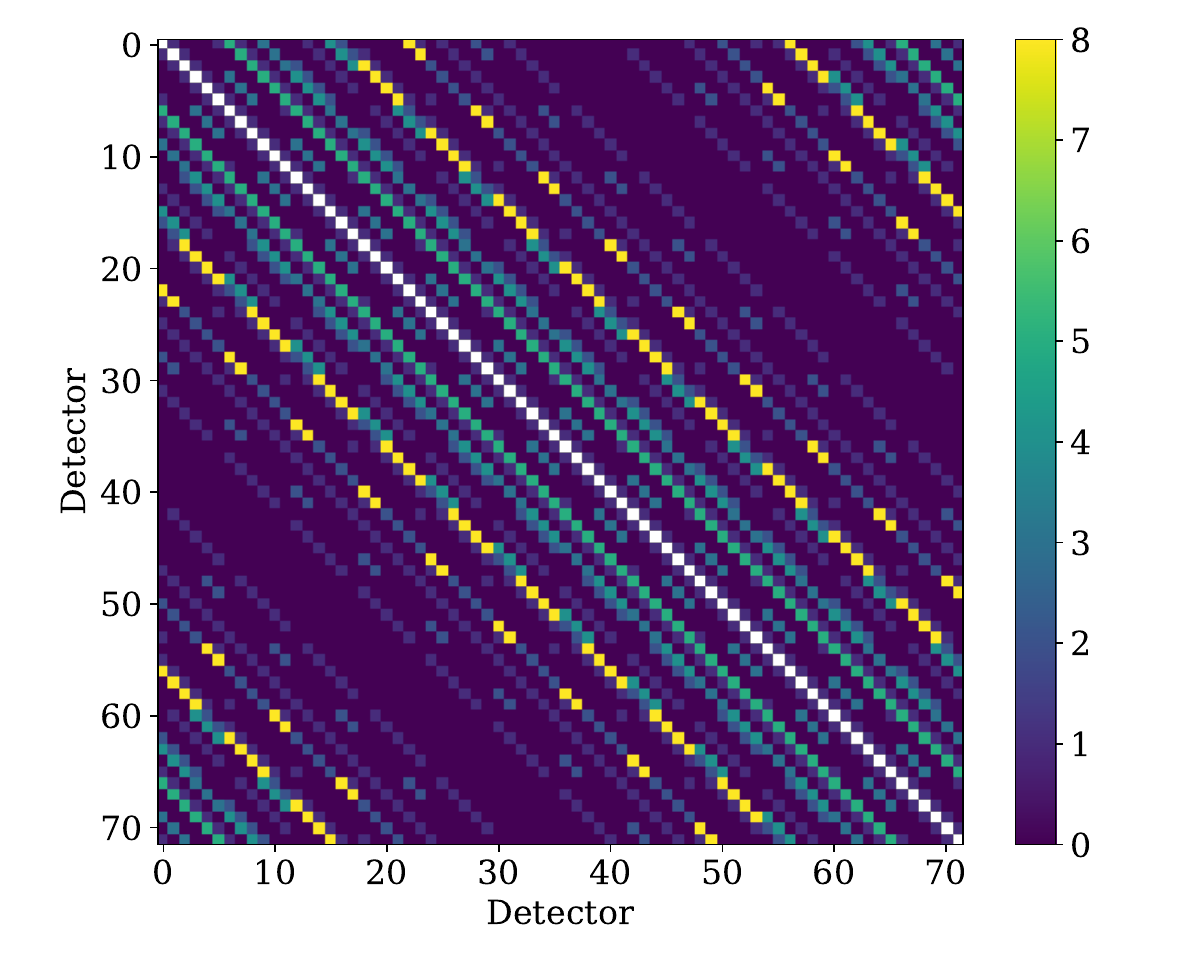}
\caption{The number of fault syndrome columns shared by each detector pair, $n_s(c_i, c_j)$ measured at the end of the first syndrome cycle, in the circuit-level decoding matrix $H_Z$ of $Z$-type errors ($X$-type stabilizers) for the gross code. The logical idle circuit analyzed here has $n/2=72$ check qubits measured at the end of each syndrome cycle, partitioning the detectors into groups of 72, and only the pairs of checks from the first group are shown. The count of shared columns of the pairs ranges from $n_s=0$ to $n_s=8$. The structure of $H_Z$ beyond the first cycle is shown in \fig{fig:z.pairs.4} (and $H_X$ with $X$-type errors in \fig{fig:x.pairs.4}), and the distribution of $n_s$ in \fig{fig:shared}. \label{fig:z.pairs}}
\end{figure}

\section{Identifying low-weight errors}\label{sec:identify}

Within the circuit-level noise model \cite{bravyi2024high}, every column of the parity-check matrices $H_X$ and $H_Z$ defines the syndrome of an isolated single circuit error (typically a few different gate errors may result in an identical syndrome and therefore one fault column). The matrix $H_X$  corresponds to $X$-type errors, detected by $Z$-type stabilizers, and $H_Z$ corresponds to $Z$-type errors, detected by $X$-type stabilizers. The decoding matrices are $(6, 35)$-sparse, with
at most 6 nonzero entries in any column and at most 35 nonzero entries in any row \cite{bravyi2024high}. An error in cycle $i$ could lead to a nonzero check in cycles $i$ and $i+1$ (but not later, and clearly not earlier). Therefore, the rows are grouped by the 72 detectors constructed from the parity change of the stabilizers with respect to the previous cycle (that we refer to simply as checks in the following), with off-diagonal elements for consecutive blocks. 
Taken together, an arbitrary pair of checks could therefore define at most $35+35=70$ columns (faults) in which one of the two corresponding checks is nonzero. For checks that appear together in one or more columns, the total number of their columns combined would be less than 70.

\begin{figure}[t!]
\centering
\includegraphics[width=0.48\textwidth]
{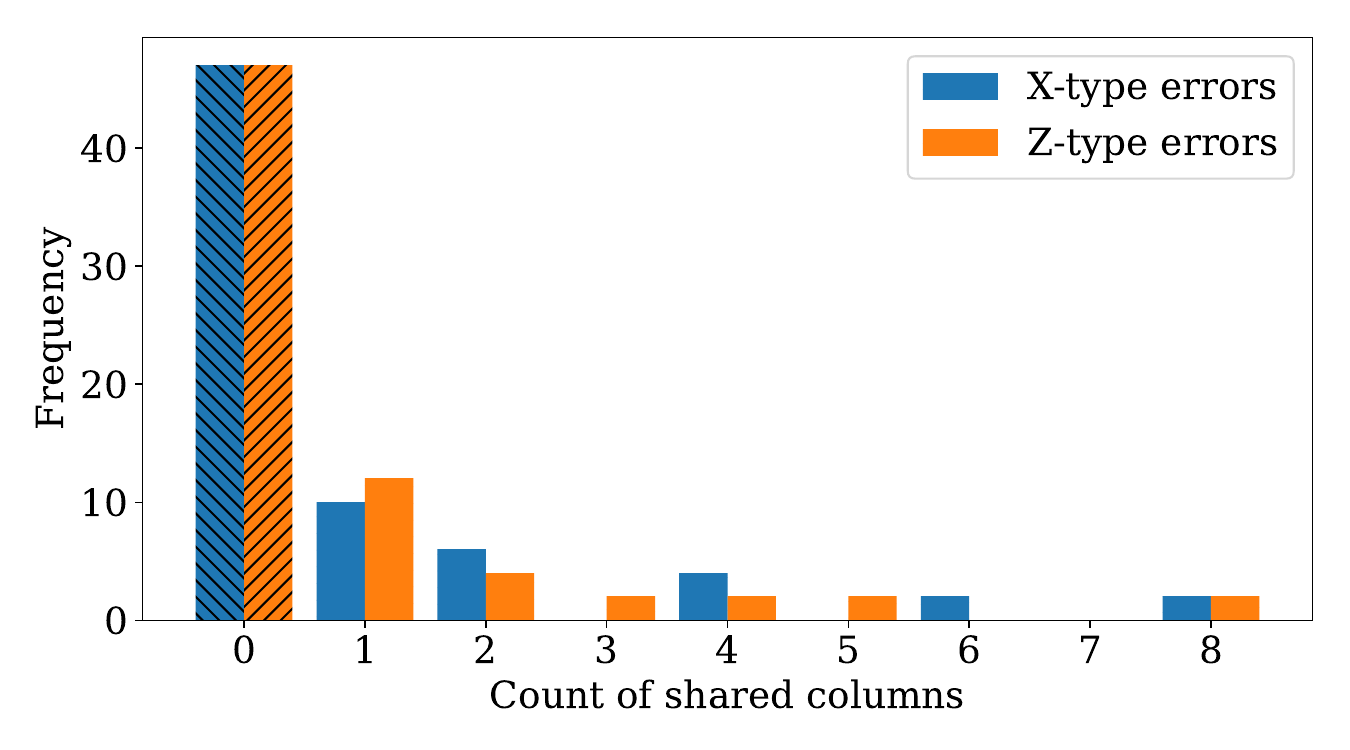}
\caption{Frequency of $n_s(c_1, c_i)$ among the 71 checks $c_i\neq c_1$ in the same syndrome cycle group with which one (arbitrary) check $c_1$ shares between zero to eight fault columns. The data corresponds to the circuit-level noise decoding matrices $H_Z$ (first group of checks, as in \fig{fig:z.pairs}) and $H_X$ (\fig{fig:x.pairs.4}, second group of checks) for the gross code with the idle logic cycle.\label{fig:shared}}
\end{figure}

Figure \ref{fig:z.pairs} shows the number of check-matrix columns shared by each pair of checks of the first group of 72 rows of $H_Z$. A very similar pattern is found in $H_X$ starting from the second syndrome cycle of the circuit, as shown in \fig{fig:x.pairs.4} in \app{sec:app}, but the syndrome circuit breaks the $X$ and $Z$ symmetry in a way that is manifested in \figs{fig:z.pairs.4}{fig:x.pairs.4}.
As can be seen in \fig{fig:shared}, every check shares a maximal number of fault columns ($n_s=8$) with two other checks, shares $n_s=5$ ($n_s=6$) columns with two other checks in the first cycle of $H_Z$ ($H_X$), and shares gradually decreasing column counts with progressively more and more checks.
We find by numerical simulations that low-weight decoding errors can be constructed from the shared columns of pairs of checks sharing $n_s=8$ columns, and we did not find similar constructions for pairs with $n_s<8$. Hence we focus on pairs for which $n_s=8$; there are 72 such pairs within a syndrome cycle. 

However, not all syndromes constructed from shared columns are found to lead to decoding errors or slow BP convergence. In order to further specify those we define a few quantities and the notation to be used below. A weight-$N$ error is the sum of $N$ distinct fault columns of the decoding matrix,
\be s\equiv s(s_1,...,s_N) = \sum_i^N s_{i}\,({\rm mod }\,2),\label{eq:error}\ee
where $s_i\neq s_j\, \forall i,j$. 
The weight of a syndrome or a fault -- not to be confused with a weight-$N$ error defined in \eq{eq:error} -- is its Hamming weight
\be w(s) \equiv |s| = \sum_j s^{(j)},\label{eq:w}\ee
$s^{(j)}$ being the value of $j$'th row bit. A somewhat less standard quantity is the number of unique checks in a set of syndromes
\be n_u(s_1,...,s_m) = \left\lvert \bigcup\{ c_i \in s_j\}\right\rvert,\ee
where the notation $c_k\in s_i$ denotes that check $c_k$ is nonzero in the syndrome $s_i$.
Using the above two definitions, the number of canceled checks for a set of syndromes is the difference between the number of unique checks and the sum syndrome weight,
\be n_c(s_1,...,s_m) = n_u(s_1,...,s_m) - w(s_1,...,s_m),\ee
effectively counting the checks that appear in an even number of the fault columns summed in the syndrome, and hence canceled in it.

Fixing an arbitrary check $c_1$, there are two pairs $(c_1, c_2)$ and $(c_1, c_{2'})$ for which $n_s(c_1, c_2)=n_s(c_1, c_{2'})=8$. The statistics corresponding to both pairs are equivalent, and we can focus on one of those, $p_0\equiv (c_1, c_2)$, for the discussion in this paragraph. Since $p_0$ defines by construction $n_s=8$ columns that the pair of checks share, it is clear that the summation of any two faults taken from these 8 columns will have both $c_1$ and $c_2$ canceled to zero. There are 28 pairs of columns that can be chosen from these combinations, pairs of syndromes $(s_1, s_2)$ obeying
\be c_1, c_2 \in s_1, s_2.\label{eq:c_1c_2}\ee
Then a second pair of checks $p_1\equiv (c_3, c_4)$ -- which might have a check coincidence with one of the checks in $(c_1, c_2)$ -- can be chosen among the remaining 71 out of 72 pairs of checks $p_i$ that share exactly $n_s(p_i)=8$ columns, and for $p_1$ again 28 combinations of the pair's shared columns can be formed in pairs of syndromes $(s_3, s_4)$ that obey
\be c_3, c_4 \in s_3, s_4,\label{eq:c_3c_4}\ee
and as described above already, 
\be n_s(c_1, c_2)=n_s(c_3, c_4)=8.\label{eq:n_sc_0c_1c_2}\ee

\begin{figure}[t!]
\centering
\includegraphics[width=0.48\textwidth]
{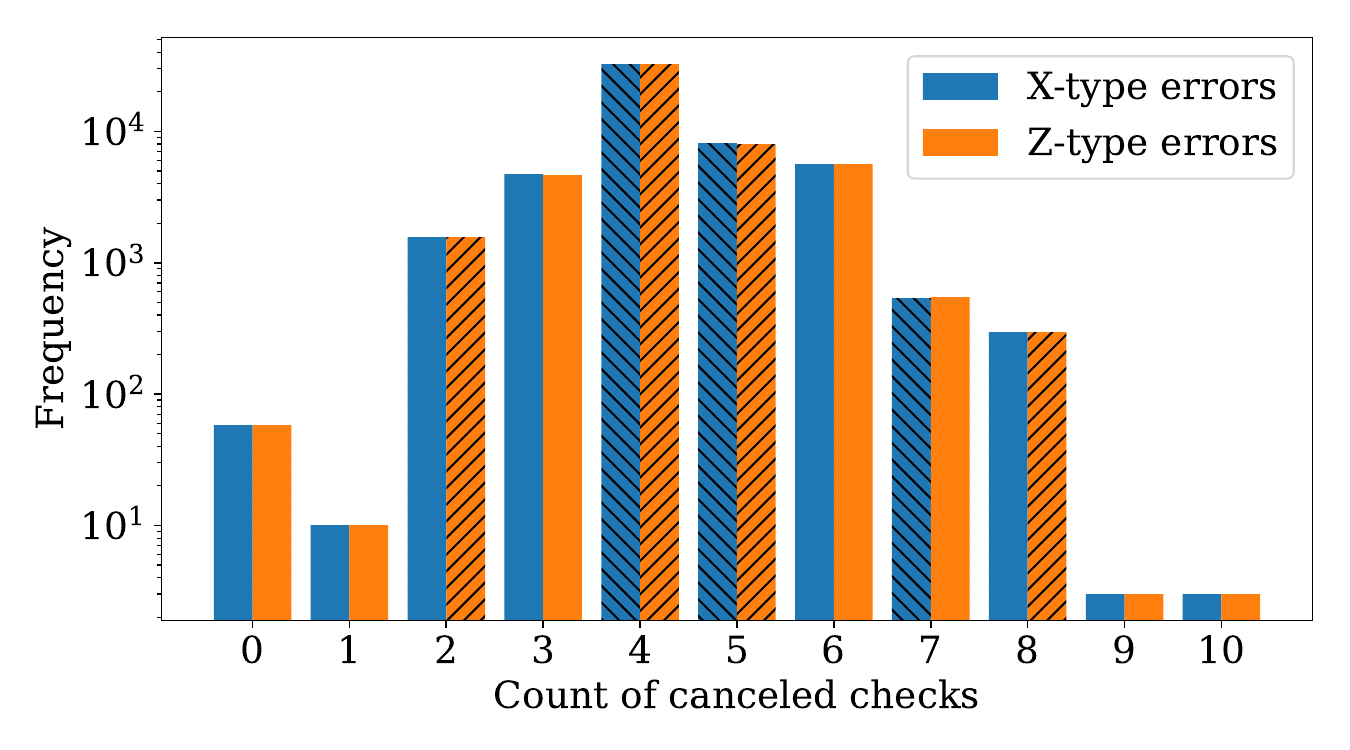}
\caption{Distribution of the number of total canceled checks ($n_c$) in the syndromes of all weight-four errors constructed by choosing any combination of two of the shared columns of the fixed pair $p_0=(c_1,c_2)$ together with any two of the shared columns of every other pair $p_i=(c_3,c_4)$ of the 71 maximal $n_s=8$ pairs in one cycle of $H_Z$ (first group of checks, as in \fig{fig:z.pairs}) and $H_X$ (second group of checks, shown in \fig{fig:x.pairs.4}).\label{fig:canceled}}
\end{figure}

With the procedure described above and the conditions in \eqss{eq:n_sc_0c_1c_2}{eq:c_3c_4} we have therefore constructed a weight-four error syndrome;
\be s \equiv \sum_{i=1}^4 s_i,\ee
(in a relatively small fraction of the cases this procedure would lead to less than four unique columns, and these errors with weight less than four can be ignored). In most resulting syndromes all four checks $\{c_1,c_2,c_3,c_4\}$ would end up canceled to zero in the combined syndrome, except in some cases where some of these checks may appear in one column chosen for the other pair.
We find that syndromes constructed in this manner have between a total of $n_c=0$ and $n_c=10$ canceled checks (that appear in exactly two or four of the four fault columns summed in the syndrome, and canceled in the total syndrome). Figure \ref{fig:canceled} shows the distribution of $n_c$ for all weight-four error syndromes (53,214 in total) constructed as described above.

\begin{figure}[t!]
\centering
\includegraphics[width=0.43\textwidth]
{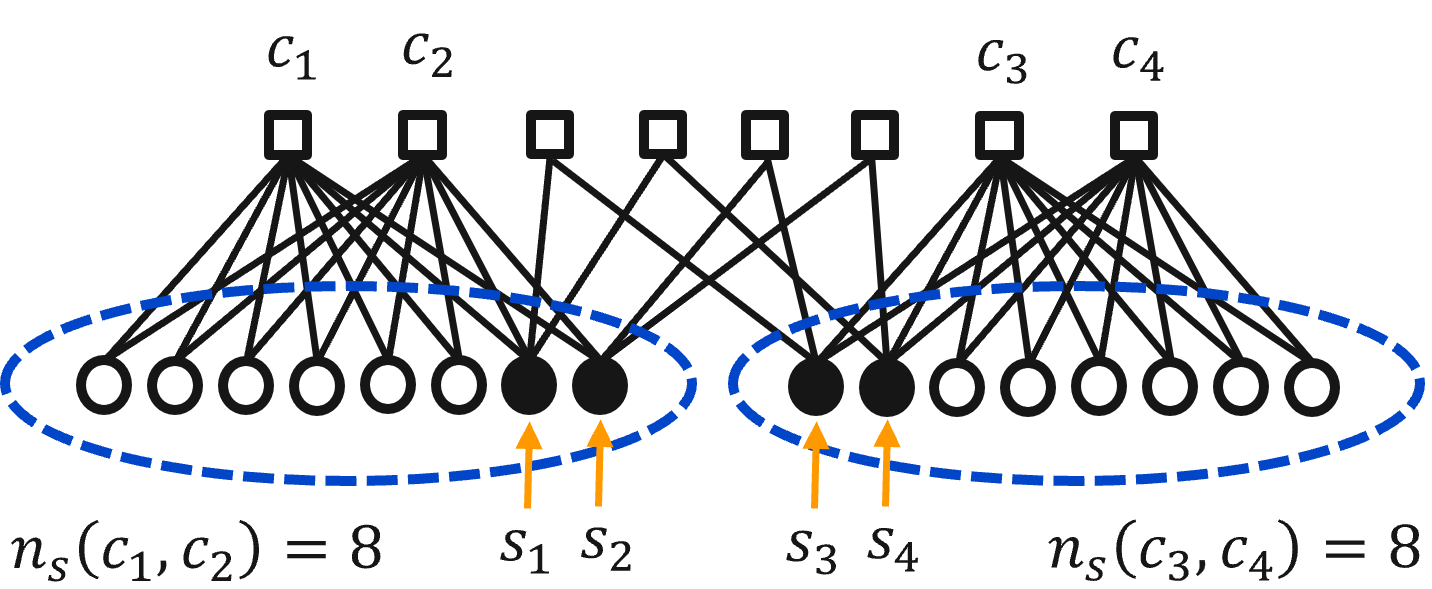}
\caption{A possible structure of part of the Tanner subgraph of weight-four errors that are slow to converge with BP iterations. Checks are drawn as squares and faults as circles, lines connect checks to their faults, and only the eight canceled checks are explicitly shown, similar to the structure in Tab.~\ref{structure_table}. \label{fig:tanner}}
\end{figure}

We can now write down the conditions allowing to identify the majority of low-weight errors; choosing shared columns of pairs of checks for which $n_s=8$, it is necessary that
\be n_c(s_1,s_2) = 2, \qquad n_c(s_3,s_4) = 2,\label{eq:condition1}\ee
while at the same time,
\be
n_c(s_1,s_2,s_3,s_4) = 8.\label{eq:condition2}\ee
These conditions imply choosing only pairs of syndromes such that exactly two checks are canceled in each pair [\eq{eq:condition1}], but such that combined together, the four syndromes add up to cancel eight checks [\eq{eq:condition2}]. The key here are the cancellations from cross-shared checks between the pairs $(s_1,s_2)$ and $(s_3,s_4)$ -- that themselves are chosen from the ``large space'' of eight shared columns for each pair.
This structure, that -- loosely speaking -- makes it apparently very ``confusing'' for BP, is shown schematically in \fig{fig:tanner} and in Tab.~\ref{structure_table}.

\begin{table}[h]
\begin{tabular}
{ |c|c|c|c| }
 \hline
 \multicolumn{2}{|c|}{Pair 0 ($p_0$)} & \multicolumn{2}{|c|}{Pair 1 ($p_1$)} \\ 
 \hline
 \, Fault 1 \,&\, Fault 2 \,&\, Fault 3 \,& \, Fault 4 \,\\ %\hline
 \,($s_1$)\,&\, ($s_2$) \, &\, ($s_3$)\, & \, ($s_4$) \,\\
 \hline \hline
 $\checksym$ & $\checksym$ & \, & \, \\
 \hline
 $\checksym$ & $\checksym$ & \, & \, \\
 \hline
 \, & \, & $\checksym$ & $\checksym$ \\
 \hline
 \, & \, & $\checksym$ & $\checksym$ \\
 \hline \hline
 $\checksym$ & \, & \, &  $\checksym$ \\
 \hline
 $\checksym$ & \, &  $\checksym$ & \,  \\
 \hline
 \, & $\checksym$ & $\checksym$ & \, \\
 \hline
 \, & $\checksym$ & \, & $\checksym$ \\
 \hline \hline
 ... & ... & ... & ... \\
 \hline
 \end{tabular}
\caption{A possible structure of the canceled checks (depicted in the rows) in the syndromes of weight-four errors that are slow to converge with BP iterations. The condition in \eq{eq:condition1} implies that within each of the pairs $(s_1,s_2)$ and $(s_3,s_4)$ exactly two checks are canceled, but put together, there are eight checks canceled between the pairs [\eq{eq:condition2}]. Each column has a maximum of six checks, so contributes eventually at most two checks (not shown) to the total error syndrome -- those are found to have a final weight between $w(s)=4$ and $w(s)=8$ [for more details see \app{sec:app}].}
\label{structure_table}
\end{table}

\section{Dynamics of low-weight errors with belief-propagation decoding}\label{sec:dynamics}

\begin{figure}[t!]
\centering
\includegraphics[width=0.48\textwidth]
{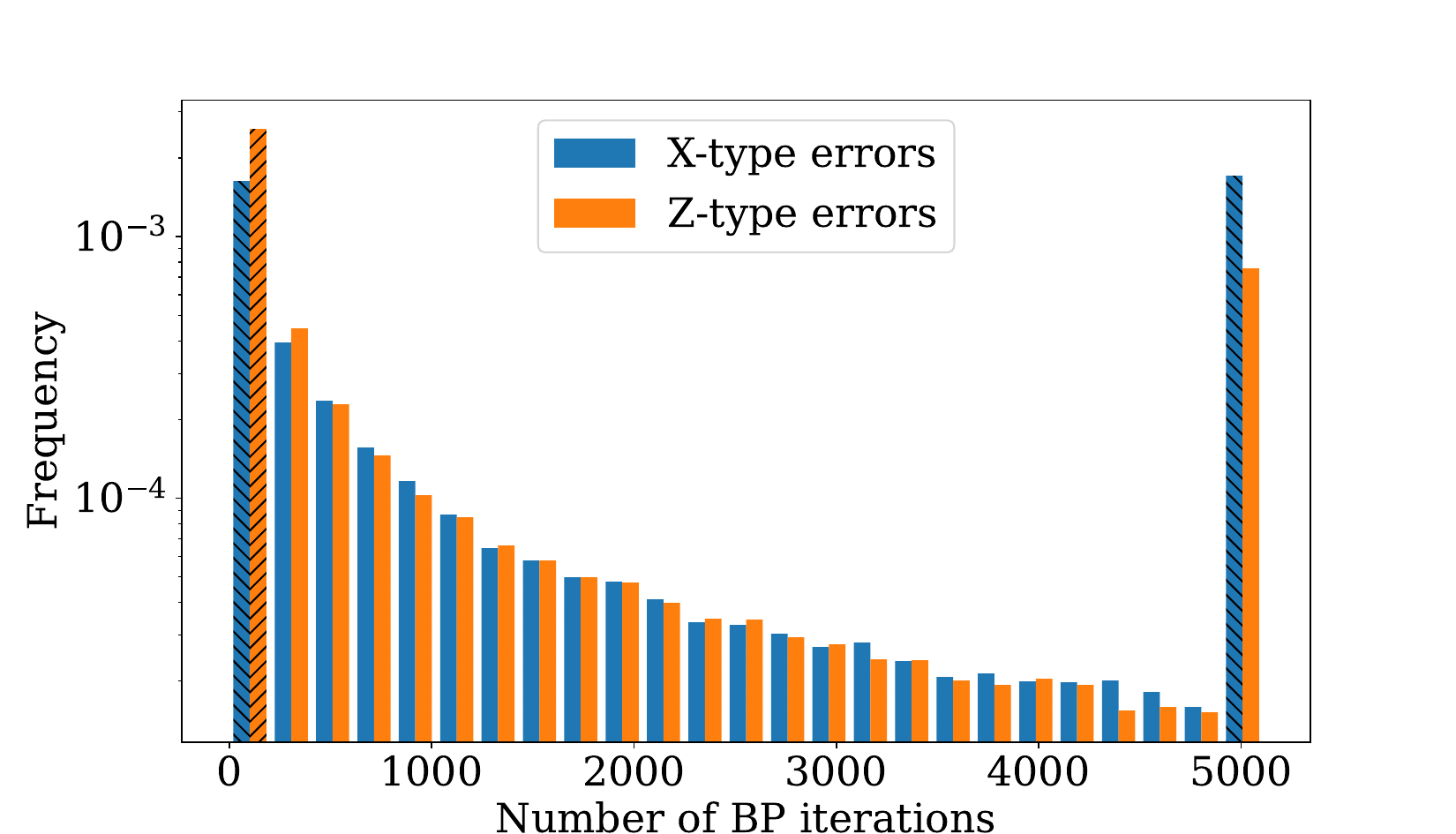}
\caption{Distribution of the total number of BP iterations in decoding of weight-four errors, selected according to the criteria in \eqss{eq:c_1c_2}{eq:condition2}, starting with the pairs of checks sharing $n_s=8$ fault columns in one cycle of $H_Z$ and $H_X$. The simulations executed 200 Relay-BP legs with a maximum of 25 iterations for each one, terminated at 5,000 itearation or with the first solution compatible with the syndrome. \label{fig:iterations}}
\end{figure}

Figure \ref{fig:iterations} shows the probability distribution of the number of BP iterations in decoding attempts of all syndromes obeying the conditions in \eqss{eq:c_1c_2}{eq:condition2} chosen for one cycle. There are 2,664 distinct such weight-four errors in one cycle of both $H_Z$ and $H_X$. In those simulations, up to 200 Relay-BP legs with a maximum of 25 iterations per leg were executed, and the first valid (syndrome-compatible) solution -- if any -- terminates the decoding (the last histogram bar contains mostly the nonconverged cases within 5,000 iterations). The simulations are performed using open-source code similar to \cite{feedforward,qec_flip_repo}, and is available at \cite{low_weight_repo}. The nature of Relay-BP with a specified distribution of memory strengths \cite{muller2025improved} including negative ones (in the interval [-0.24, 0.66]), which are being randomized at every leg, introduces a stochastic sampling of the BP dynamics for each error. Each fault was simulated 50 times in order to obtain convergence statistics. It is clear from the figure that there is a significant tail of errors that converge either very slowly or not at all within the cutoff, leading with a high probability to a logical decoding error. This slow convergence should be contrasted with that of weight-four error syndromes taken from the $n_s=8$ pairs without conditioning on the canceled checks ({\it i.e.}, from the population depicted in \fig{fig:canceled}), requiring just 5 (7) iterations in the mean for $Z$ ($X$) errors to converge.

We note that that relaxing either one of the two conditions in \eqss{eq:condition1}{eq:condition2}, to include a total of $n_c=6$ and $n_c=7$ canceled checks for weight-four error syndromes, and to construct those from column pairs with $n_c=4$ canceled checks, we find that some further syndromes with a slow-down in the BP decoding, but it appears that those typically converge eventually. Errors with a total of $n_c=9,\,10$ canceled checks are observed to be either trivial (in the syndrome and their logical action) or easy for the decoder.

\begin{figure}[t!]
\centering
\includegraphics[width=0.48\textwidth]
{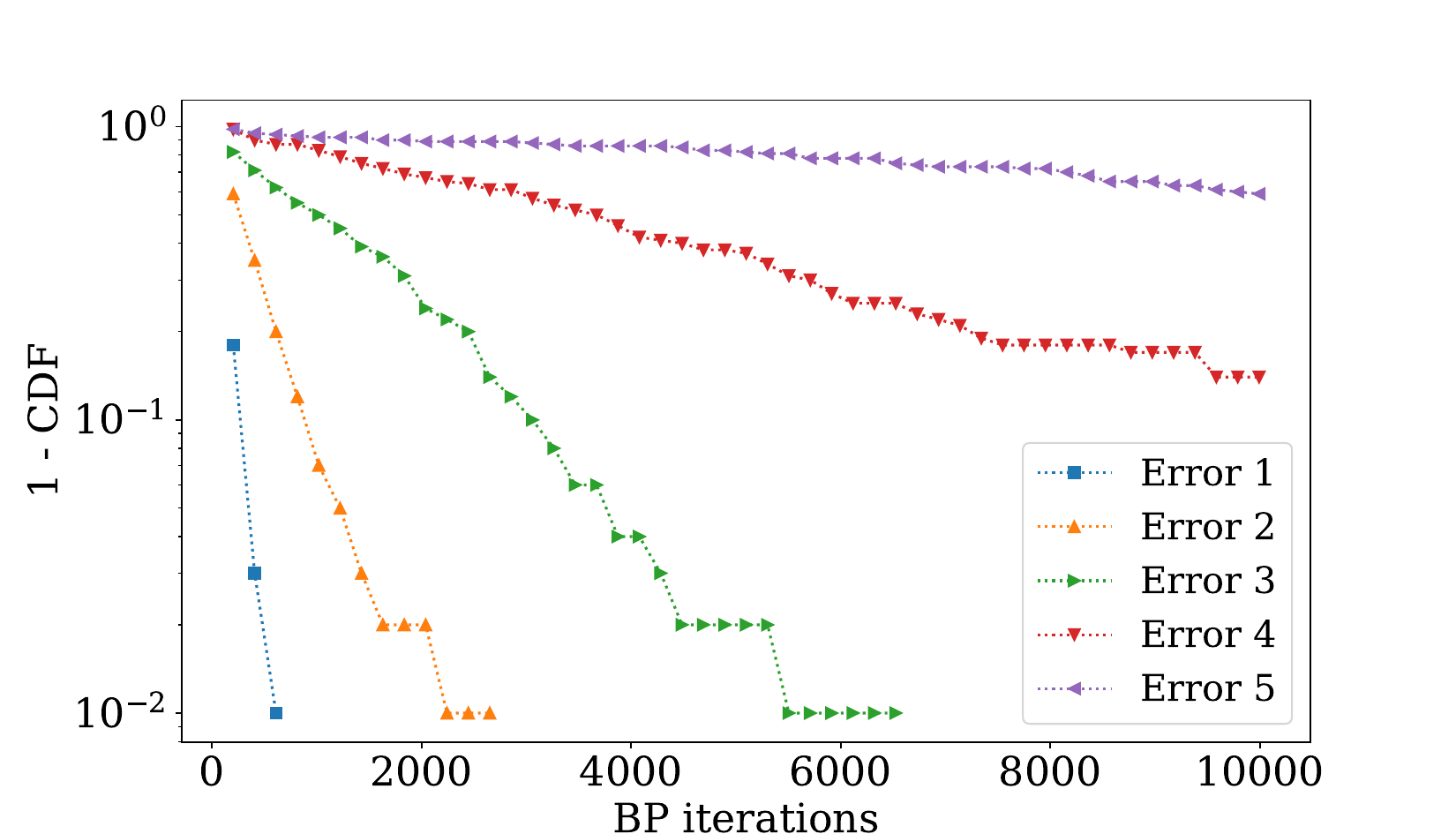}
\caption{Distribution of the total number of BP iterations for a few exemplary $Z$-type weight-four errors, simulated with up to 200 Relay-BP legs with a maximum of 50 iterations per leg (and hence a cutoff of 10,000 iterations). The plot shows $1-P(N \le n)$ with $P$ the binned cumulative distribution function for the number of BP iterations, and 100 simulations were repeated for each error. Due to the log-scale plotting, the final $P=1$ values of each curve were omitted, and the finite sampling explains the plateaus at low values. The approximately straight lines are plausibly consistent with exponential statistics with a very broad range of rates between the errors. \label{fig:z.some.iterations}}
\end{figure}

To elaborate more on the BP dynamics hiding behind the statistics in \fig{fig:iterations}, we further simulate each error 100 times, with 200 BP legs and a maximum of 50 iterations per leg (reaching up to 10,000 BP iterations). Using those simulations we can study the convergence dynamics of each individual weight-four error. Some curves of the cumulative probability distribution sampled for exemplary errors are plotted in \fig{fig:z.some.iterations}. The BP convergence rates for specific errors are consistent with an exponential escape distribution, similar to well-known mechanisms of thermal activation (where Relay-BP provides the stochastic kicks by construction), or leak from a chaotic phase-space domain (where the decoding matrix environment could plausibly induce those dynamics, see below).

The very broad range for the BP convergence rate -- from a few tens to tens of thousands of BP iterations or more -- suggests that conditions \eqss{eq:condition1}{eq:condition2} still do not tell the full story of the convergence of BP iterations on the constructed low-weight errors. We have studied the relevance of different further parameters that can be calculated for the syndromes of the low-weight errors. In \app{sec:app} we point to the weight of the final syndrome as having some correlation with the BP convergence rate in the case of the $X$-type errors of the first syndrome cycle, but not in general.
There remains an unidentified mechanism that determines the distribution of BP convergence for each error. An analysis of the distributions of the number of BP iterations executed to convergence shows several distinct possibilities, compatible either with a power-law density, or an exponential density, and in some cases apparently requiring more than a single parameter [see \app{sec:app}]. 

\begin{figure}[t!]
\centering
\includegraphics[width=0.48\textwidth]
{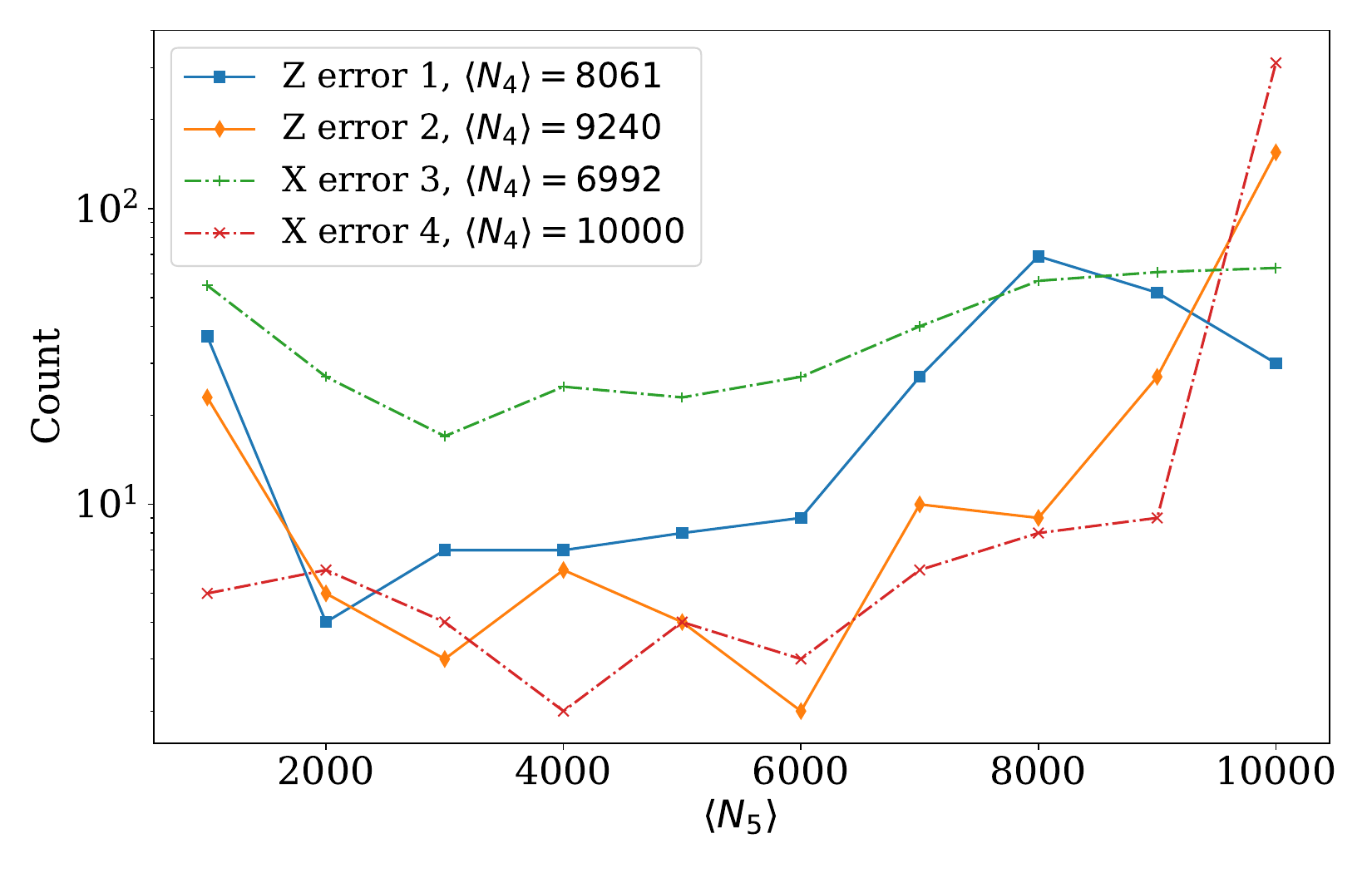}
\caption{Binned counts of weight-five errors in 10 bins of 1,000 BP iterations-wide, for errors constructed from each of a few exemplary weight-four errors, simulated with parameters as in \fig{fig:z.some.iterations}. $\langle N_5\rangle$ denotes the mean number of BP iterations from 100 independent simulations for each error of the few hundreds of the weight-five errors constructed from one weight-four error whose mean number of BP iterations is $\langle N_4\rangle$. Adding a fifth fault induces very large changes with a complex spread to the convergence rate as compared with the weight-four error, serving as an indication that the rate is determined by a many-body dynamical process. \label{fig:randomizations}}
\end{figure}

Although there may well be more algebraic structure characterizing the errors than discovered in this work, the above observations appear to provide some evidence that the convergence rate of individual errors may not be determined by one simple parameter (or more) of the error syndrome, but rather by the entire neighborhood of the errors in the Tanner graph represented by the decoding matrix.
To support this, we start with the weight-four errors identified above and add to their syndrome a fifth fault column, to study the ensuing weight-five decoding dynamics. To make this empirical experiment nontrivial, we choose the fifth column to have at least one check in common with the checks of the original four faults composing the low-weight error, so that the fifth fault is part of their local neighborhood in the decoding matrix's Tanner graph. We simulate 100 times the decoding of every one of a few hundreds of weight-five error constructed in this way from each basic weight-four error, and obtain the mean BP convergence rate. Figure \ref{fig:randomizations} shows the count of weight-five errors binned into bins of 1,000 iterations-wide, corresponding to each of a few exemplary weight-four errors. It can be seen that the weight-five errors have a spread of BP convergence rates across two orders of magnitude with different behaviors. The observed complexity appears incompatible with a simple characterization and is a possible indication that the decoding dynamics are truly many-body in the involved checks and error nodes of the Tanner graph.

\section{Resolving low-weight decoding errors}\label{sec:resolve}

Perhaps the most immediate approach to combat low-weight errors would be to amend the decoding matrices in a way making the convergence of BP efficient in those situations (this is arguably simpler than changing the decoding algorithm or trying to modify the code itself). 
As a starting point, we find that directly adding all weight-four error syndromes studied above to the decoding matrices appears perfectly effective; BP converges within tens of iterations on all such syndromes with no logical errors. However, there is a relatively large number of these errors and increasing the decoding matrices incurs penalties;  it might have a negative effect on the decoding of generic errors, and it may be limited by hardware constraints in real-time decoding implementations. In particular, it would lead to an increase in the decoding time since the decoding matrices are larger and redundant in the fault columns. 

\begin{figure}[t!]
\centering
\includegraphics[width=0.48\textwidth]
{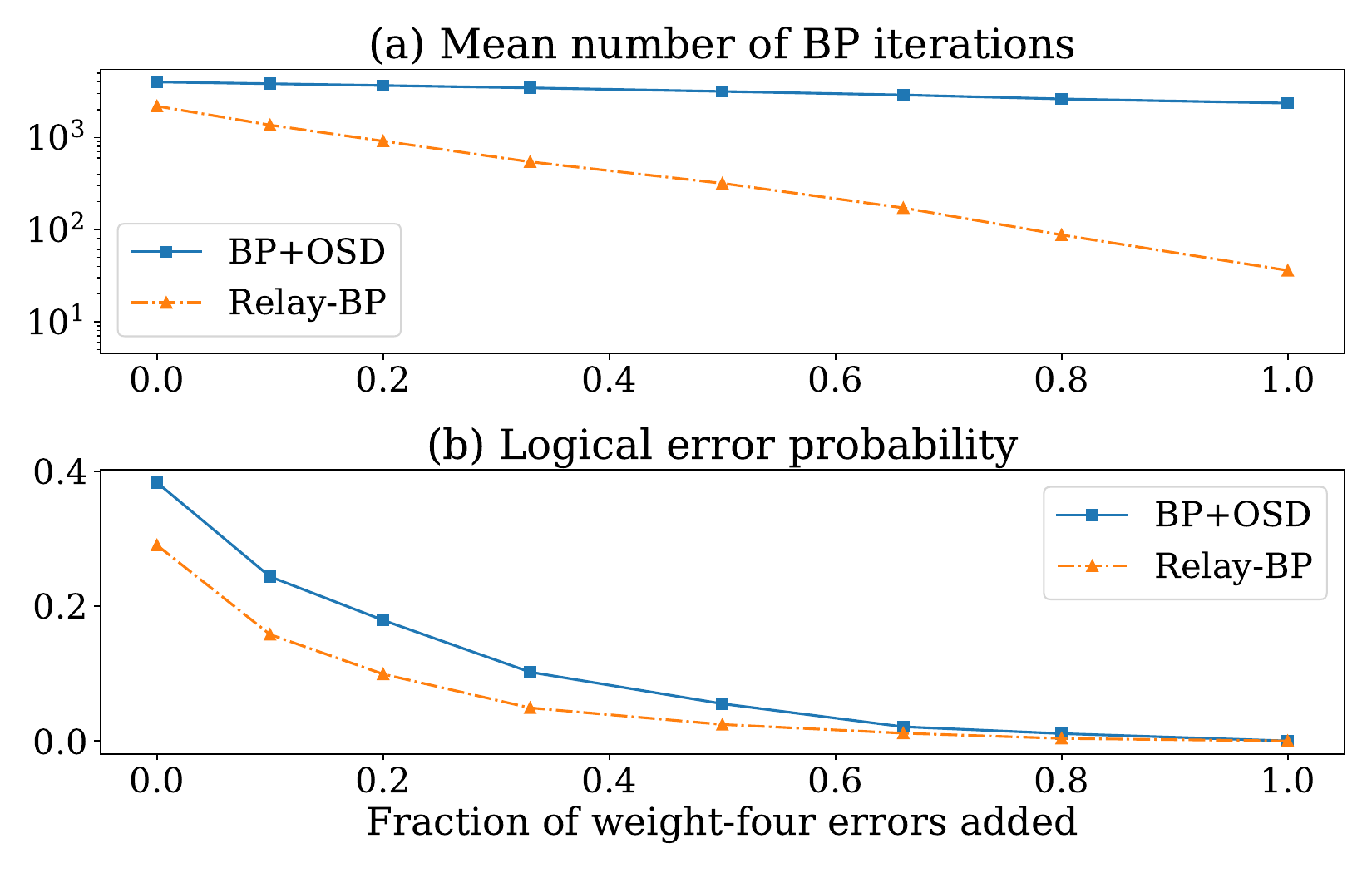}
\caption{(a) The mean number of BP iterations and (b) the logical error probability for decoding of the weight-four errors of $X$-type in \fig{fig:iterations}, as a function of the fraction of those errors that are added as independent columns into the decoding matrix. Two curves are shown, for decoding with BP with a maximum of 5,000 iterations and OSD as a secondary decoder, and for using Relay-BP. The logical error probability drops exponentially with both decoders, but Relay-BP is significantly more efficient, as could be expected.\label{fig:amended}}
\end{figure}

For example, in one syndrome cycle we find that there are 1,164 distinct column pairs corresponding to \eq{eq:condition1}, from which 2,664 distinct weight-four errors can be constructed to obey both \eqss{eq:condition1}{eq:condition2}. These weight-four errors can in turn be broken into 3,888 distinct pairs. We have tested the possibility to improve BP convergence when including in the decoding matrices new columns composed of sums of pairs of faults (in different combinations), but the results were not optimal.
Instead, we quantify the effectiveness of a simple alternative stochastic approach that could allow limiting the inflation in decoding-matrix columns. In \fig{fig:amended} we consider adding to the $X$-type decoding matrix the syndromes of a random fraction out of all weight-four errors (as above, of the first cycle).
Figure \ref{fig:amended} shows the reduction in the mean number of BP iterations and in the logical error rate in decoding with similar parameters as in \fig{fig:iterations}, as a function of the fraction of those errors that are added as independent columns into the decoding matrix. The decay curves are very close to exponential and this could indicate a direction for an optimization of a possible tradeoff.

The question of optimizing the tradeoff between costs and gains in a realistic setup depends on details that are beyond the scope of the present study. In particular, given the results of \seq{sec:dynamics}, it seems important to do such an optimization accounting for the full error spectrum of a stochastic decoding simulation. This would be most interesting not with the gross code studied here, for which the low-weight error floor appears to be quite low \cite{muller2025improved,beverland2025fail} and therefore perhaps not very relevant, but with more complex syndrome cycles crucial for fault-tolerant computations and where the error floor is perhaps a more urgent limitation \cite{yoder2025tour}. The code used for the simulations of the current work is available as open source \cite{low_weight_repo}, and could possibly be adapted to those goals.

In such setups, beyond an offline amending of the decoding matrices as considered here, more complex approaches can be explored. If patterns in the error syndromes that are suggestive of the hard errors can be identified ({\it e.g.}, particular check combinations), the decoding could be redirected to one (or more) dedicated decoders, tailored to the suspected hard errors, with extended decoding matrices, adapted Relay-BP parameters (such as the effective ``temperature'' or disorder allowing to explore the phase space faster), or even a change of algorithm in the more general case. Alternatively, it could be the irregularly slow BP convergence (if it can be identified) that would trigger an activation of dedicated decoders, and in setups where enough parallelization is feasible \cite{seelam2026reference}, this could be attempted even without any indication at all.

To conclude, applying the approach described here to identify low-weight errors, discover more algebraic structure characterizing the errors and exploring ideas for their decoding in relevant syndrome cycles, could possibly contribute to reducing logical failures in cycles that are of high importance for fault-tolerant quantum computing. 

%More generally, if a parameter that correlates strongly with the BP convergence rate for low-weight errors can be identified from the syndromes, it would provide a valuable criterion. However, a simple criterion might not be explicitly identified if the convergence is determined by a large number of interactions of the fault columns and other columns with their neighborhood in the Tanner graph, as the results in \seq{sec:dynamics} may indicate. In this case, brute-force numerical decoding simulations can still be used to possibly further screen relevant individual errors according to their convergence rates, unless the number of candidate low-weight errors from recognized criteria are already too large (which might imply that the decoding-matrix extension suggested here would not be viable as well).

{\it Acknowledgments.} We thank Andrew Cross, Tristan M\"uller, Dekel Meirom, Tsafrir Armon, Liran Shirizly, Thilo Maurer, Patrick Rall, Lev Bishop, Anirudh Krishna and Kevin Krsulich for very helpful feedback.

\appendix

\section{More details on the decoding dynamics}\label{sec:app}

\begin{figure}
\centering
\includegraphics[width=0.48\textwidth]
{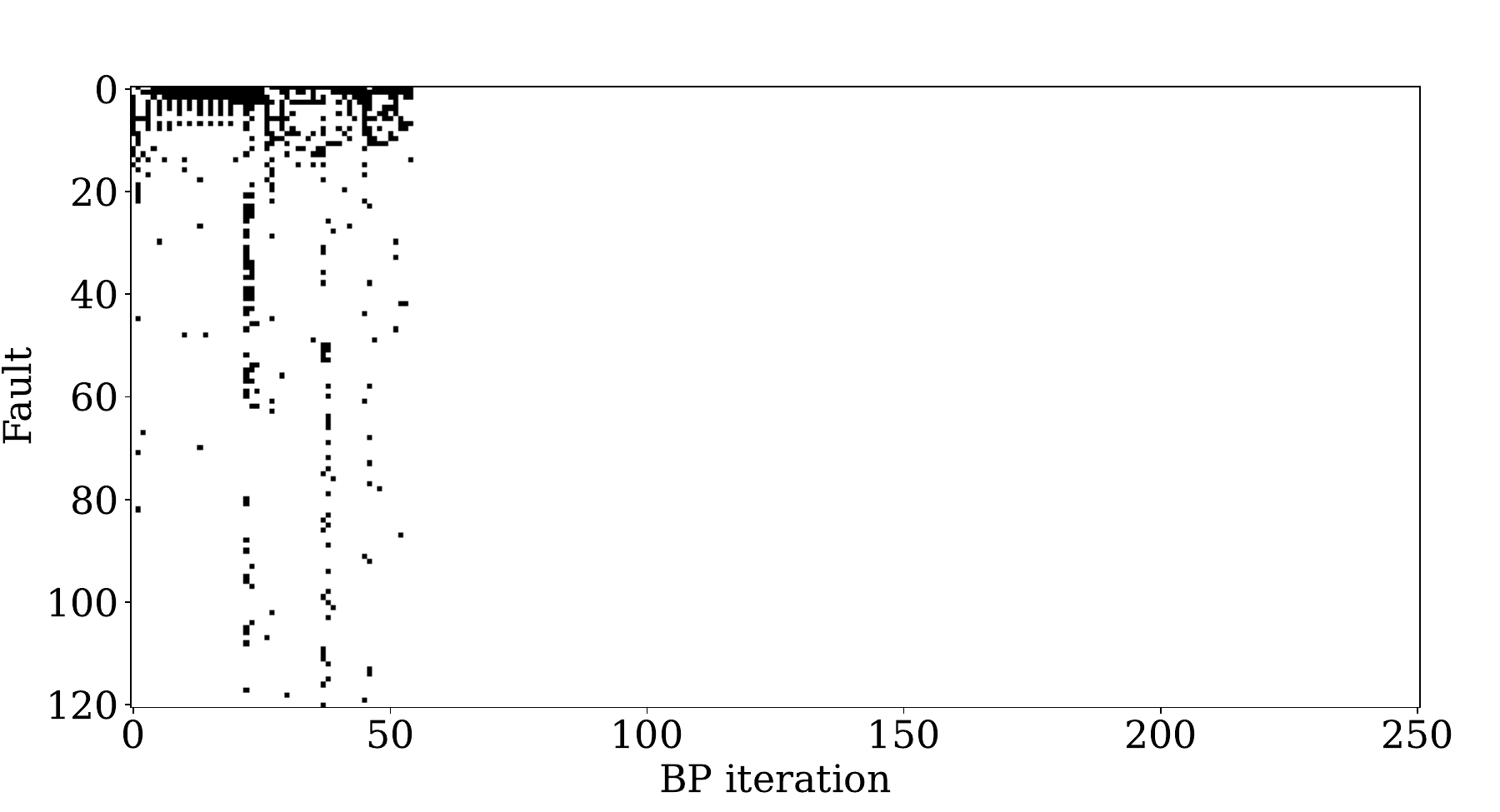}
\caption{The ``hard decision'' bit of part of the faults (variable nodes) of the decoding matrix, similarly to \fig{fig:fault}, at the end of the BP iterations of a relatively rapidly converging Relay-BP execution. Here, a low-weight error syndrome converges within 54 iterations of the first, warm-up relay leg without positive damping 
to a correct decoding. The faults are arbitrarily ordered by their total brightness (and truncated beyond the top 121 ones). Despite many faults ``blipping'' during the decoding, the convergence to the correct few ones is relatively quick, and the execution is terminated.
\label{fig:success}}
\end{figure}

The Relay-BP simulations in this work are carried out with standard parameters taken from those given in \cite{muller2025improved}, with the memory weights randomly chosen with a uniform distribution from the interval $[-0.24,0.66]$.

For a contrasting comparison with \fig{fig:fault}, a low-weight error that converges relatively quickly is shown in \fig{fig:success}. Within the first (warm-up) leg of BP iterations the correct error is converged and the iterations terminated.

\begin{figure}[t!]
\centering
\includegraphics[width=0.5\textwidth]
{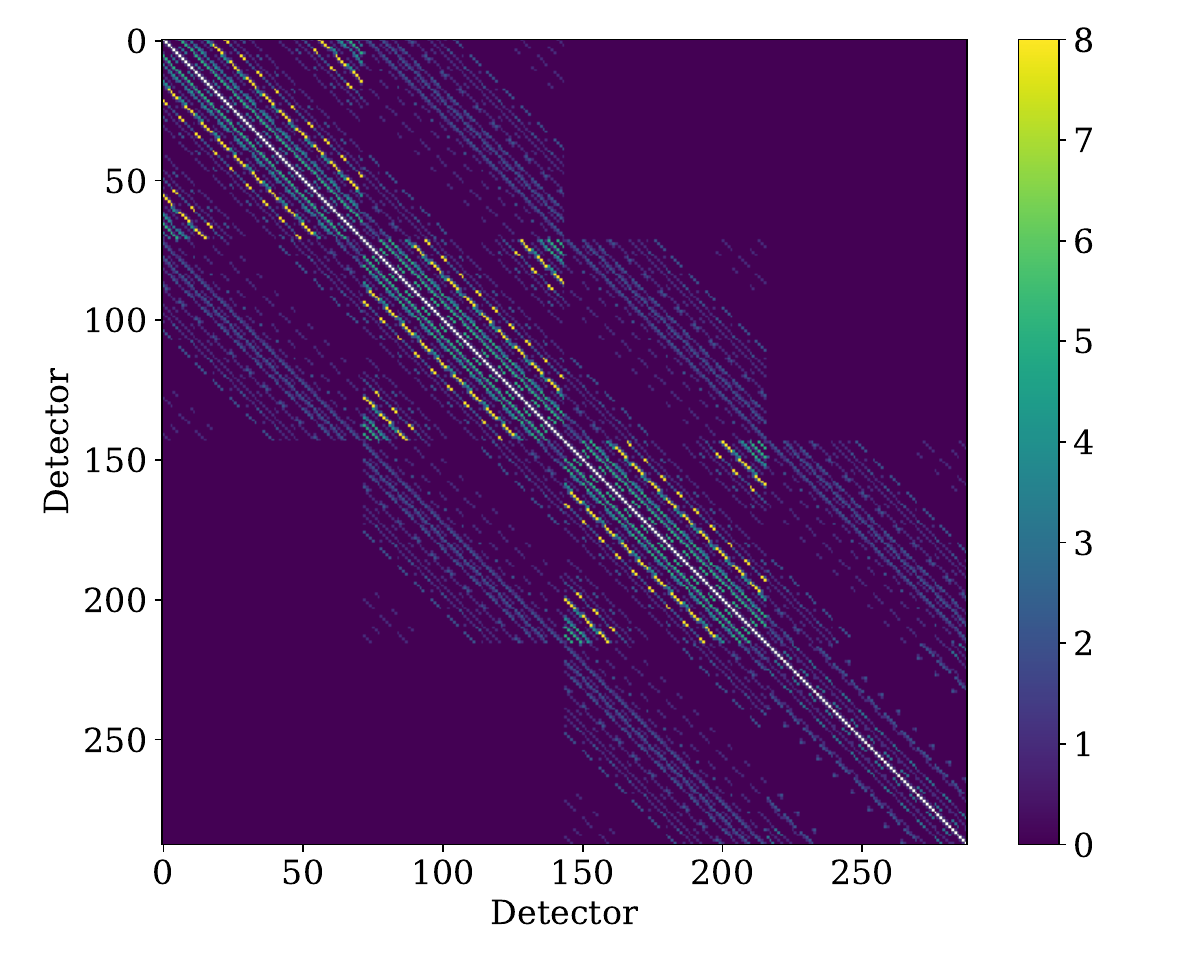}
\caption{The number $n_s$ of fault syndrome columns shared by each detector pair in the circuit-level noise decoding matrix $H_Z$ of $Z$-type errors for the gross code with the idle cycle, extending \fig{fig:z.pairs}. The syndrome measurement circuit here has three noisy cycles and one final noiseless cycle. The syndrome cycle circuit breaks the symmetry between $X$- and $Z$-type errors and the cycle-translation invariance, as can be seen by comparison with the structure of $H_X$ shown in \fig{fig:x.pairs.4}.\label{fig:z.pairs.4}}
\end{figure}

\begin{figure}
\centering
\includegraphics[width=0.5\textwidth]
{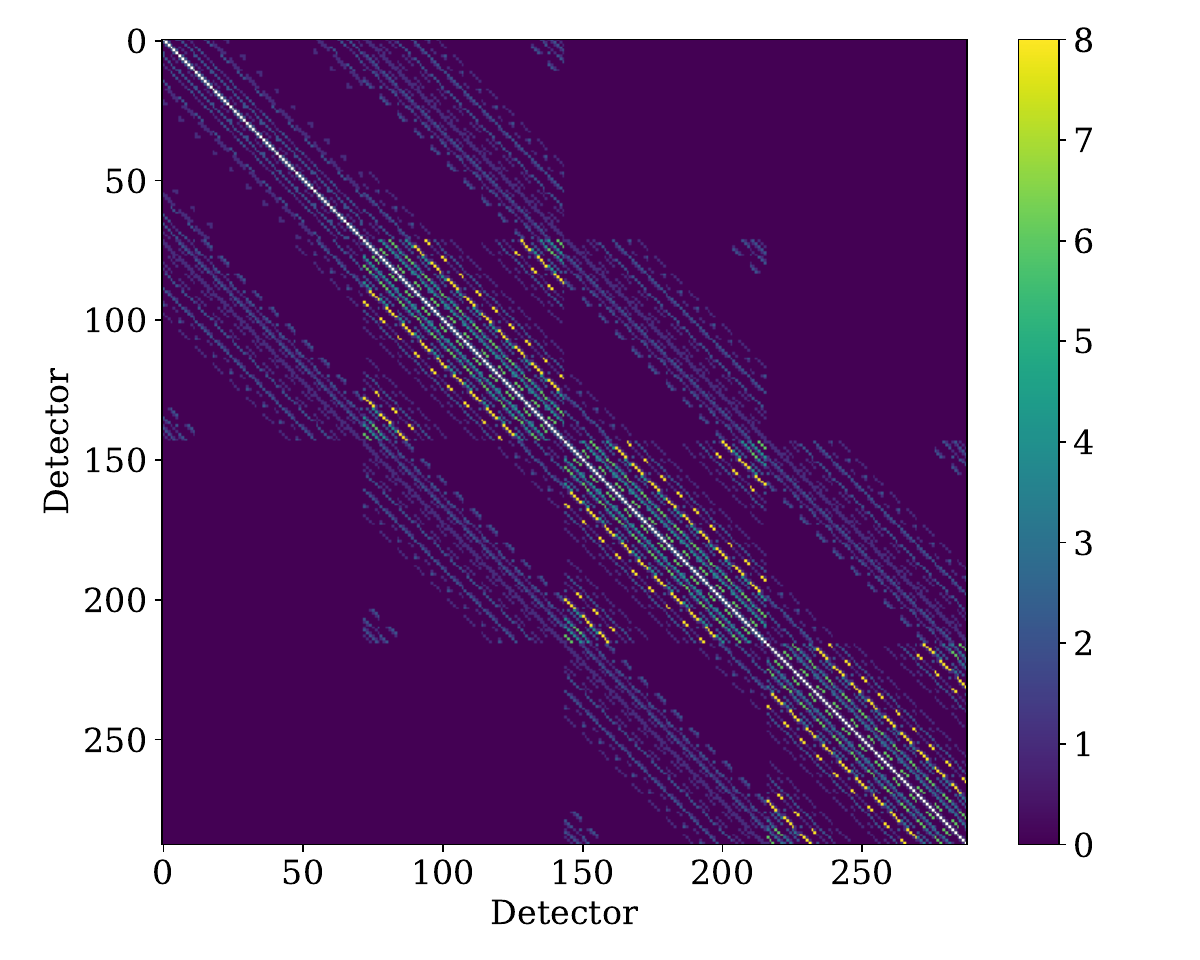}
\caption{The number $n_s$ of fault syndrome columns shared by each detector pair in the circuit-level noise decoding matrix $H_X$ of $X$-type errors ($Z$-type stabilizers) for the gross code with the idle cycle, similar to \fig{fig:z.pairs.4}. \label{fig:x.pairs.4}}
\end{figure}

\begin{figure}
\centering
\includegraphics[width=0.46\textwidth]
{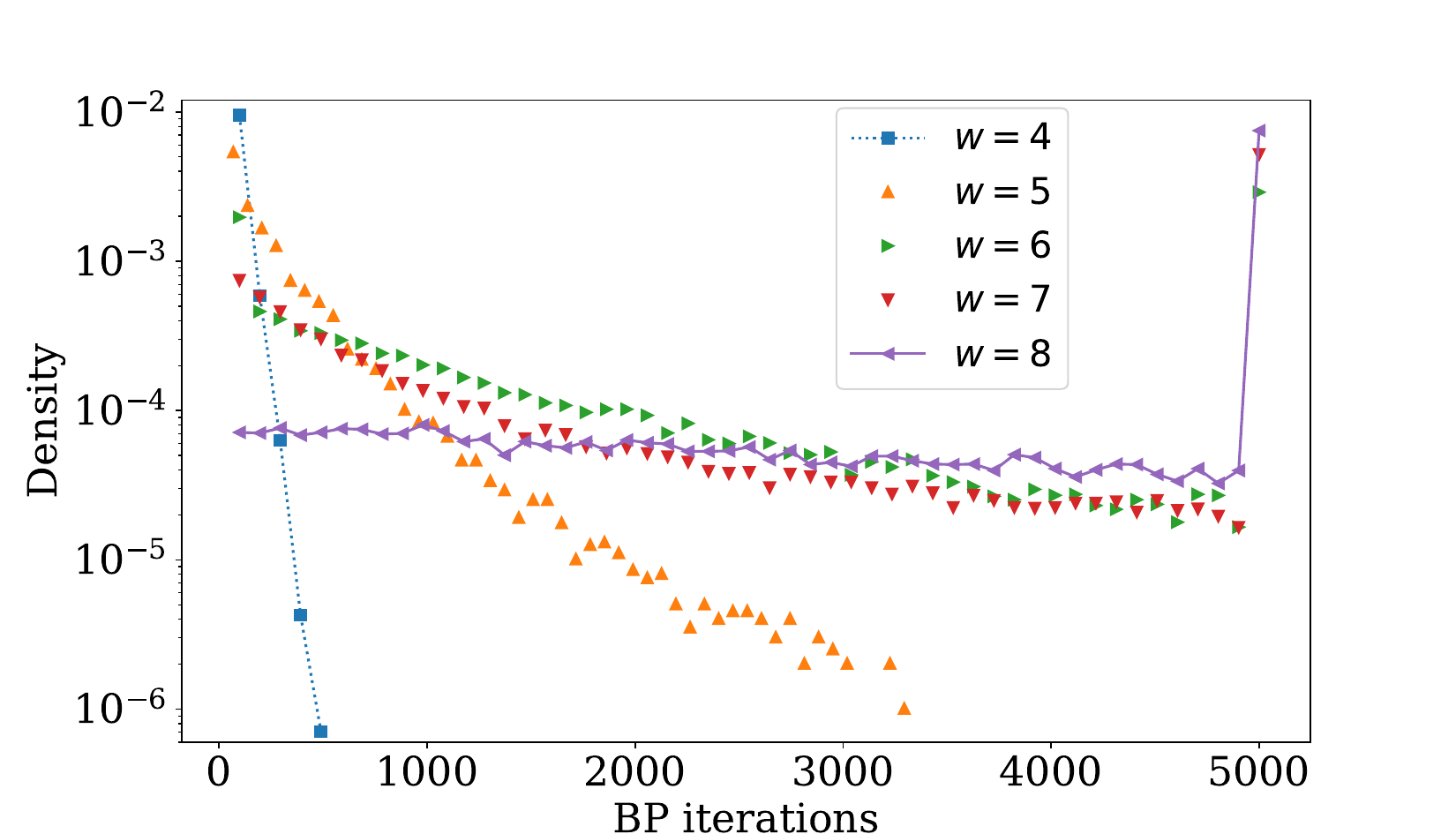}
\caption{Distribution of the total number of BP iterations as in \fig{fig:iterations} for $X$-type errors, separated according to the value of $w(s)$ defined in \eq{eq:w}. The lines drawn for $w=4$ and $w=8$ are guides to the eye. \label{fig:x.iterations}}
\end{figure}

\begin{figure}
\centering
\includegraphics[width=0.46\textwidth]
{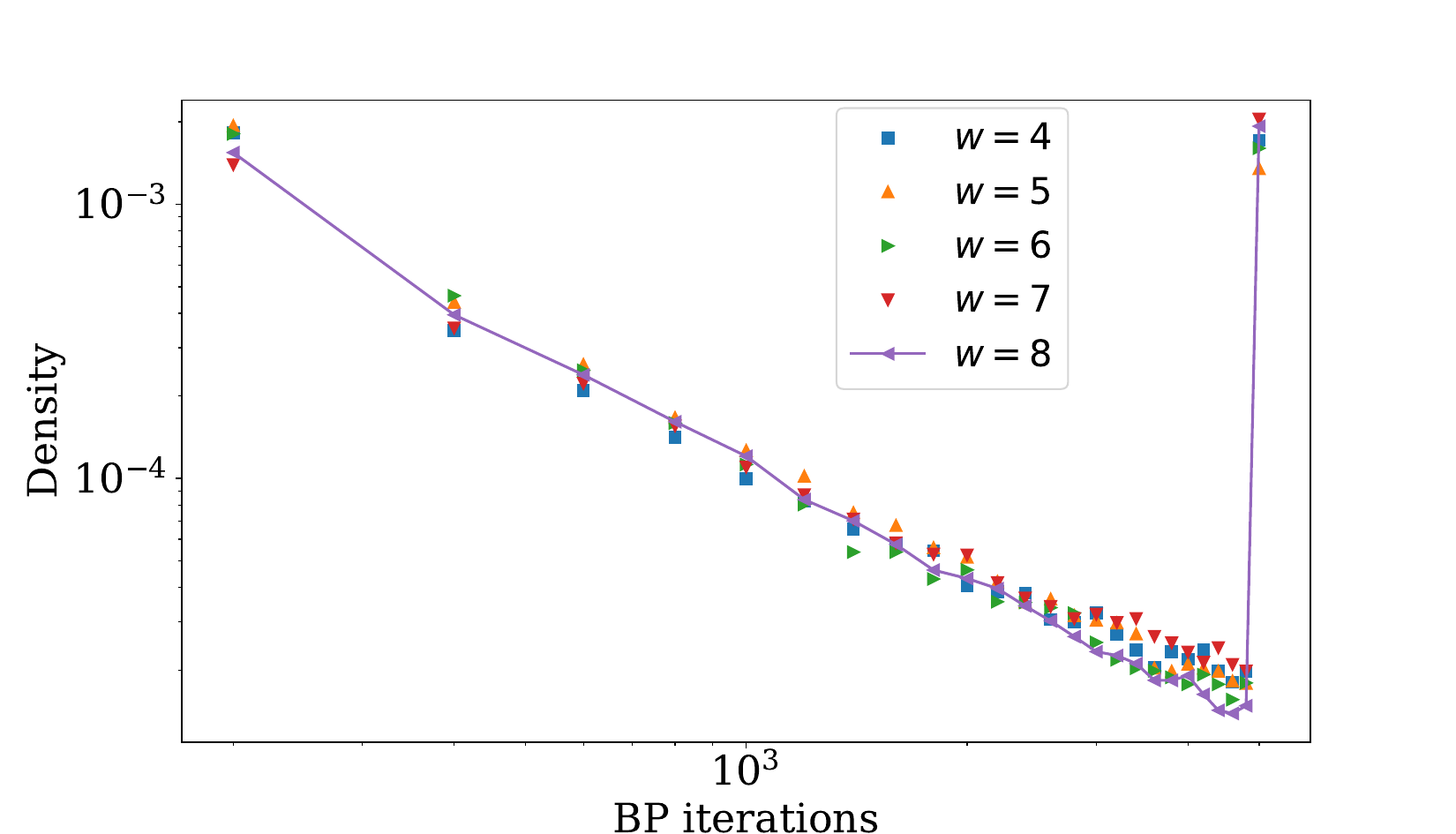}
\caption{Distribution of the total number of BP iterations as in \fig{fig:iterations} for $Z$-type errors, separated according to the value of the error syndrome weight $w$ [\eq{eq:w}]. In this plot the iterations axis and the density are plotted in log scale.\label{fig:z.iterations}}
\end{figure}

As described in the text, \fig{fig:z.pairs} shows the count of shared columns between only checks that are both in the first measured cycle. To illustrate the full structure of shared columns, we present in \figs{fig:z.pairs.4}{fig:x.pairs.4} the corresponding extension to the four cycles of a syndrome measurement circuit composed of three noisy cycles and one final ideal (noiseless) cycle. The breaking of symmetry between the $X$ and $Z$ matrices is visible.

Despite -- as described in \seq{sec:dynamics} -- the lack of an exact handle on the iteration distributions of individual errors, we might get some clue from examining the overall convergence statistics. We find a partial further characterization of the BP convergence rates by looking at the weight of the final syndrome (the final number of checks remaining in the syndrome fed into the decoder), which lies in the range  $4 \le w(s) \le 8$.
Figure \ref{fig:x.iterations} shows the distribution of BP iterations of $X$-type errors from the same decoding simulations as in \fig{fig:iterations}, separated according to the value of $w(s)$. The number of BP iterations across all errors at lower values $w=4$ and $w=5$ are roughly consistent with exponential distributions. For higher values the distribution does not seem to be characterized by one parameter. A similar analysis for the $Z$-type faults (in the first cycle) shows a less informative picture, where in \fig{fig:z.iterations} the corresponding BP iterations seem to be somewhat close to a power-law distribution (as seen by a plot in log-log scale), with relatively small differences between different $w$ values.

\bibliography{citations}

\end{document}